\documentclass[10pt]{article}
\usepackage{epsfig}
\begin{document}
\begin{center}
The Study of Two-dimensional Polytropic Stars

\bigskip
Sanchari De$^a$, Sutapa Ghosh$^b$  and Somenath Chakrabarty$^a$$^\dagger$

\medskip
$^a$Department of Physics, Visva-Bharati, Santiniketan-731235, 
India\\
$^\dagger$somenath.chakrabarty@visva-bharati.ac.in\\
$^b$Department of Physics, Barasat Govt. College, 
Barasat 700124, India
\end{center}

\begin{center}
Abstract
\end{center}

\noindent 
A formalism has been developed to investigate some of the
gross properties of hypothetical two-dimensional polytropic stars.
We strongly believe that the present investigation will throw some light on
the structure and also on some of the gross properties of such 
two-dimensional self-gravitating stellar
objects. The other interesting part of the work is to study the
polytropic equation state of two-dimensional dense stellar matter.

\section{Introduction}
Since the temperature inside the main-sequence stars are quite high,
the thermal energy of the hydrogen ions are large enough to
overcome the inter-ionic Coulomb barrier. As a result the conversion of 
hydrogen to helium is continuously going on inside such stars. For the 
stars of mass very close to the sun, i.e., for the lower
main-sequence stars, the thermonuclear fusion processes are
taking place through $p-p$-chain reactions. Whereas for the massive
or upper main-sequence stars, 
the hydrogen to helium conversion process is called the CNO-cycle. 
To study the structure  and also some of 
the gross properties of main-sequence stars the solution of the well 
known Lane-Emden equation \cite{ST} is used. This equation is
essentially obtained from the hydrostatic equilibrium condition inside
the star. For the 
physically acceptable values of various parameters of the star, in 
particular the mass and the radius, the numerical
solution of Lane-Emden equation is used. To obtain this
equation from the hydrostatic equilibrium condition, the pressure and 
the density of stellar matter are assumed to be related by some equation 
of state, 
known as the polytropic equation of state. The general form of this
equation is given by 
\begin{equation}
P=K\rho^{\Gamma}
\end{equation}
where $P$ and $\rho$ are respectively the pressure and the density of
stellar matter, $K$ is a constant which  depends on the nature of stellar 
matter 
and the constant $\Gamma$ is called the polytropic index. In stellar
astrophysics, instead of $\Gamma$, for the sake of convenience it is 
expressed as 
\begin{equation}
\Gamma=1+\frac{1}{n}
\end{equation}
and $n$ is then treated as the polytropic index. In the usual main-sequence 
stellar model $n=3/2$ for extreme non-relativistic
case and $n=3$ for the ultra-relativistic situation.

In this article we have
developed a formalism to study the 
structure and some of the gross properties of two-dimensional hypothetical 
main-sequence stars. To the best of our knowledge such studies have
not been reported in the past. 
In \cite{NDIM}, the Lane-Emden equation has been studied in
$N$-dimension. But the theme of that work was to obtain various types
of mathematical solutions for Lane-Emden equation in higher
dimension, without going into the physics of the problem.
Our present work is essentially a two-dimensional  extension
of conventional polytropic model for main-sequence stars, discussed in
many standard text book on astrophysics \cite{ST,RQ}.
\section{Basic Formalism}
For the two-dimensional stellar objects, we redefine the pressure as the 
force per unit length. Then it is very easy to show that the hydro-static 
equilibrium condition is given by
\begin{equation}
dP+g(r)\rho(r)dr=0
\end{equation}
Now in two-dimension, since the gravitational potential $\sim
\ln(r)$, the gravitational force per unit mass may be expressed 
in the form
\begin{equation}
g(r)=\frac{G}{r}\int_0^r 2\pi\rho (r^{\prime})r^{\prime}dr^{\prime}
\end{equation}
Combing these two equations (eqn.(3) and eqn.(4)), we can rewrite the 
hydro-static equilibrium condition in the following form
\begin{equation}
\frac{r}{\rho}\frac{dP}{dr}+G\int_0^r2\pi\rho r^{\prime}dr^{\prime}=0
\end{equation}
On differentiating throughout by $r$ and after rearranging some 
of the terms, we have
\begin{equation}
\frac{1}{r}\frac{d}{gr}\left (\frac{r}{\rho(r)}\frac{dP}{dr}\right )+
2\pi G\rho(r)=0
\end{equation}
Now using the polytropic equation of state (eqn.(1)) and replacing the 
redial coordinate $r$ by the usual scaled radial parameter $x$,
defined by $r=ax$, where $a$ is an unknown constant, we have
\begin{equation}
\frac{(n+1)K}{a^2nx}\frac{d}{dx}\left (x\rho^{\frac{1}{n}+1}
\frac{d\rho}{dx}\right )+2\pi G\rho=0
\end{equation}
Further, using the techniques followed in the conventional three-dimensional 
scenario, we put
\begin{equation}
\rho=\rho_c\theta^n
\end{equation}
where $\rho_c$ is the central density and $\theta$ is some dimensionless 
variable. Then the above equation (eqn.(7)) can be rewritten in the
following form
\begin{equation}
\frac{1}{x}\frac{d}{dx}\left ( x \frac{d\theta}{dx}\right )=-\theta^n
\end{equation}
with 
\begin{equation}
a=\left [\frac{(n+1)K\rho_c^{\frac{1}{n}-1}}{2\pi G}\right ]^{\frac{1}{2}}
\end{equation}
Eqn.(9) is the two-dimensional version of Lane-Emden equation.
Whereas the conventional three dimensional form is given by
\begin{equation}
\frac{1}{x^2}\frac{d}{dx}\left ( x^2 \frac{d\theta}{dx}\right )=-\theta^n
\end{equation}
Of course with different expression for the constant $a$ \cite{ST}.
To solve the Lane-emden equation (eqn.(9)), which is a second order 
differential equation, we need two initial
conditions or one initial and one boundary condition.
At the centre, $\rho=\rho_c$, therefore $\theta=1$, which is the
maximum value of $\theta$ and treated as the initial condition.
The surface of the star is given by $\rho=0$, which gives $\theta=0$,
which is the boundary condition. The
corresponding radial coordinate will be the radius of the star.
We have seen that only for $n=0$ and $n=1$, the Lane-Emden  equation 
in two-dimension can be solved analytically. For $n=0$, eqn.(9) becomes
\begin{equation}
\frac{1}{x}\frac{d}{dx}\left ( x \frac{d\theta}{dx}\right ) =-1
\end{equation}
Using the initial condition. the solution of this equation is given by 
\begin{equation}
\theta=1-\frac{x^2}{4} 
\end{equation}
Since $a=\infty$ for $n=0$, the radius of the star also becomes infinitely 
large.
The above solution is therefore not physically acceptable (this is also 
true for the usual
three dimensional case). Now for $n=1$, the Lane-Emden equation
in two-dimension can be expressed in the form
\begin{equation}
\frac{d^2\theta}{dx^2}+\frac{1}{x}\frac{d\theta}{dx}+\theta=0
\end{equation}
which is the well known Bessel differential equation of order  zero. 
The solution
is given by $\theta(x)=AJ_0(x)$, where $A$ is a constant \cite{AS}.
Since at $x=0$, $\theta=1$ and $J_0(0)=1$, the constant $A=1$. Now
for $n=1$, the radius of the object is found to be independent of
central density $\rho_c$ ($a$ becomes independent of $\rho_c$).
Therefore the analytical solution of Lane-Emden equation with $n=1$
is also unphysical (this is also true for three
dimensional situation).  For general $n$-values, 
the Lane-Emden equation is therefore solved numerically 
with the initial and boundary conditions mentioned above.

The mass of the star is obtained from the integral
\begin{equation}
M=\int_0^R 2\pi rdr\rho(r)
\end{equation}
Expressing $\rho$ in terms of the variable $\theta$ and  $r$ in terms of $x$ 
in the Lane-Emden equation (eqn.(9)), we finally have 
\begin{equation}
M=-2\pi a^2\rho_c x_s\frac{d\theta}{dx}\mid_{x=x_s}
\end{equation}
where $x_s$ is the surface value of the scaled radius parameter. The surface
term of the derivative is obtained from the numerical solution of
Lane-Emden equation. The actual radius of the star is then given by 
\begin{equation}
R=sx_s
\end{equation}
Now to obtain mass-radius relation for such hypothetical
stellar object, we consider two different scenarios for the internal
structure of the object. We first assume that both the kinetic pressure 
and the thermal energy density of the stellar matter are coming from
the baryonic part only, i.e., from the hydrogen ions. Then following 
some standard text book on
statistical mechanics (see for example \cite{LL}), it can very easily be 
shown that for the
non-relativistic situation, the energy density and kinetic pressure
of the matter are given by
\begin{equation}
\epsilon=\frac{g_B}{4\pi m_p\hbar^2}\int_0^\infty p^3 f(p)dp
\end{equation}
and
\begin{equation}
P=\frac{g_B}{4\pi m_p\hbar^2}\int_0^\infty p^3 f(p)dp
\end{equation}
respectively. Where $g_B$ is the spin degeneracy of baryons, $f(p)$ is the 
Fermi distribution and $m_p$ is the baryon mass. Therefore in this
scenario, the polytropic equation of state may be expressed as $P=K\rho$, 
with $\rho=\epsilon/c^2$, the equivalent mass density. Similarly for the
ultra-relativistic case when the baryon mass is neglected, we have
\begin{equation}
\epsilon=\frac{g_Bc}{2\pi \hbar^2}\int_0^\infty p^2 f(p)dp
\end{equation}
and
\begin{equation}
P=\frac{g_Bc}{4\pi \hbar^2}\int_0^\infty p^2 f(p)dp
\end{equation}
The polytropic form of equation of state will therefore be of the
same type as we have written for the non-relativistic case, of course
with different $K$. If we now compare the equation of state with the
standard form of polytropic equation, the index $n$ will be $\infty$. 
As a consequence, mass
of the star $M\propto \rho_c^{1/n}$ becomes independent of central
density and the radius of the star $R\propto \rho_c^{(1-n)/2n}$ will
become $R\propto \rho_c^{-1/2}$. 

On the other hand if we assume that the matter is a
mixture of proton and electron, then for the non-relativistic case,
the baryon mass has to be replaced by $m_B=\mu m_p$, where $\mu$ is a
number, called mean molecular weight and is $0.5$ for the case of
fully ionized hydrogen gas. Of course the
ultra-relativistic expressions will not change, only $g_B$ will be $4$
instead of $2$ because both electron and proton are fermion.

Next we consider a white dwarf like star in two dimension. The
degeneracy pressure of the electron gas makes the star stable against
gravitational collapse. On the other hand the mass of the object is
coming from the massive baryonic part, which are assumed to be in static
condition, therefore does not contribute in pressure. For such a
stellar object, the mass density is given by
\begin{equation}
\rho=n_em_p=\frac{m_p}{2\pi\hbar^2}p_{F_e},
\end{equation}
where $n_e$ is the electron surface density, given by
\begin{equation}
n_e=\frac{p_{F_e}}{2\pi \hbar^2},
\end{equation}
$m_p$ is the baryon mass and $p_{F_e}$ is the electron Fermi
momentum. Following \cite{LL}, the degeneracy pressure for electron
gas in the non-relativistic and relativistic scenarios are given by
\begin{equation}
P=\frac{1}{8\pi \hbar^2 m_e}p_{F_e}^4
\end{equation}
and
\begin{equation}
P=\frac{c}{6\pi\hbar^2}p_{F_e}^3
\end{equation}
respectively. Here we have taken the spin degeneracy for electron
$g_e=2$. Hence we can write down the polytropic equation of states in
the non-relativistic and ultra-relativistic cases in the form
\begin{equation}
P=K\rho ^2 ~~{\rm{and}}~~ P=K^\prime \rho^{3/2} ~~
{\rm{respectively}}
\end{equation}
Hence $n=1$ for the non-relativistic case and $n=2$ for the
ultra-relativistic situation. Then it can very easily be shown that
for non-relativistic scenario, the mass of the star $M\propto
\rho_c$ and the radius becomes independent of central density, which
is physically unacceptable. On the other hand for $n=2$ case, 
$M\propto \rho_c^{1/2}$ and the radius $R\propto \rho_c^{-1/4}$.
Hence the mass-radius relation is $MR^2=$ constant.

Whereas for the conventional white dwarf scenario, $n=3/2$ for the
non-relativistic case and $n=3$ for the ultra-relativistic case. If
we consider these values for two-dimensional case, 
the mass-radius relations for non-relativistic and ultra-relativistic 
scenarios are given by
$MR^4=$ constant and $MR=$ constant respectively.
\section{Conclusion}
In this work we have solved an hypothetical problem associated with
two-dimensional polytropic stars. We have also investigated the gross
properties of such stellar objects in an elaborate manner,
assuming different types of internal structures. 

\end{document}